\documentclass[12pt,letterpaper]{article}

\title{\bf Linking Covariant and Canonical General Relativity via Local Observers}
\author{\bf Steffen Gielen\\[.5em]
{\sl \small Perimeter Institute for Theoretical Physics} \\[-.3em]
{\sl \small  Waterloo ON, N2L 2Y5, Canada} \\
\small  \texttt{sgielen@perimeterinstitute.ca} 
 \and
{\bf Derek K.\ \!Wise} \\[.5em]
{\sl \small Institute for Theoretical Physics III} \\[-.3em]
{\sl \small Universit\"at Erlangen--N\"urnberg} \\[-.3em]
{\sl \small Staudtstr.\ \!7/B2,\! 91054 Erlangen,\! Germany} \\
\small \texttt{derek.wise@gravity.fau.de} 
}

\usepackage{amssymb,amsmath}
\usepackage[
    colorlinks,%
    linkcolor=blue,citecolor=red,urlcolor=blue,
]{hyperref}
\usepackage[all,knot]{xy}
\usepackage{tikz}
\usepackage{graphicx}
\usepackage{wrapfig}

\def\barr{\begin{array}}
\def\earr{\end{array}}
\def\ben{\begin{equation}}
\def\een{\end{equation}}
\def\bena{\begin{eqnarray}}
\def\eena{\end{eqnarray}}

\newcommand{\Rs}{\mathfrak{R}}  
\newcommand{\OMs}{{\bf \Omega}} 

\newcommand{\R}{{\mathbb R}}

\def\stackto #1 { \, {\stackrel{#1}{\longrightarrow}}\, }
\def\stackTo #1 { {\stackrel{#1}{\Longrightarrow}} }

\newcommand{\SO}{{\rm SO}}
\newcommand{\so}{\mathfrak{so}}

\newcommand{\define}[1]{{\bf #1}}

\newcommand{\we}{\wedge}
\renewcommand{\L}{\pounds} 

\newcommand{\hepth}[1]{\href{http://arxiv.org/abs/hep-th/#1}{arXiv:hep-th/#1}}

\newcommand{\physics}[1]{\href{http://arxiv.org/abs/physics/#1}{arXiv:physics/#1}}

\newcommand{\grqc}[1]{\href{http://arxiv.org/abs/gr-qc/#1}{arXiv:gr-qc/#1}}

\newcommand{\arxiv}[1]{\href{http://arxiv.org/abs/#1/}{arXiv:#1}}

\newcommand{\webpage}[1]{{\color{blue}}\url{#1}{\color{blue}}}

\begin{document}

\date{March 30, 2012}

\maketitle

\thispagestyle{empty}

\begin{abstract}
Hamiltonian gravity, relying on arbitrary choices of `space,' can obscure spacetime symmetries.  We present an alternative, manifestly spacetime covariant formulation that nonetheless distinguishes between `spatial' and `temporal' variables.  The key is viewing dynamical fields from the perspective of a {\em field of observers}---a unit timelike vector field that also transforms under local Lorentz transformations.  On one hand, all fields are spacetime fields, covariant under spacetime symmeties.  On the other, when the observer field is normal to a spatial foliation, the fields automatically fall into Hamiltonian form, recovering the Ashtekar formulation.   
We argue this provides a bridge between Ashtekar variables and covariant phase space methods.  We also outline a framework where the `space of observers' is fundamental, and spacetime geometry itself may be observer-dependent. 



\end{abstract}

\vfill
\begin{center}
{Essay written for the Gravity Research Foundation \\
2012 Awards for Essays on Gravitation.}
\end{center}
\section*{Nonlocal consequences of canonical gravity}

General relativity teaches us space and time are not independent, but inseparably entangled in a unified {\em spacetime}.   Nevertheless, standard procedure in canonical gravity is to temporarily disregard this lesson, foliating spacetime into spacelike level sets of some time function. This gives an initial value formulation of general relativity \cite{adm}, with many uses in classical and quantum gravity. 

But this approach, depending on an arbitrary, unobservable time function, has strange physical consequences.  While the spacetime picture of gravity is described by {\em local} equations, the foliation constrains {\em global} spacetime geometry and topology.   A well-posed initial value formulation demands global hyperbolicity, which in turn implies constant spatial topology \cite{geroch,bernal-sanchez}.  This leaves many interesting spacetimes, including even anti-de~Sitter space, with no `dynamical' description. 

Here we present an alternative, {\em fully local} description of gravitational dynamics, based on the notion of a {\it field of observers}.  This is useful for a geometric understanding of Lorentz symmetry in canonical gravity, for relating geometrodynamics with connection dynamics, for linking canonical and covariant phase spaces, and for various possible extensions of general relativity.

\section*{Observer fields}

In Minkowski space $\R^{3,1}$, an observer with velocity $y$ in hyperbolic space $H^3$ has a global notion of `space,' namely the subspace $\R^3_y\subset \R^{3,1}$ orthogonal to $y$:
\[
\xy
(0,0)*{\includegraphics{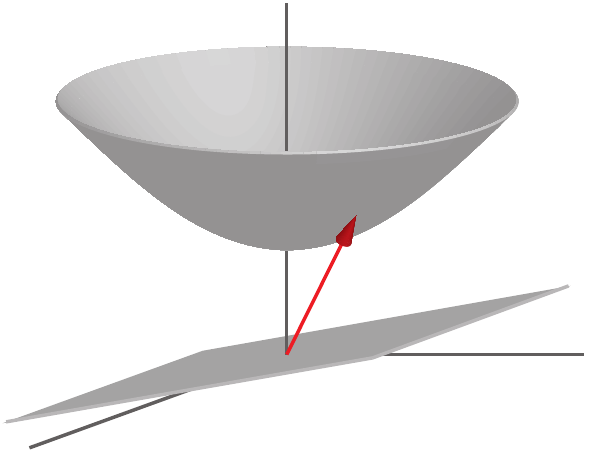}};
(26,12)*{H^3};
(38,12)*{\txt{\sf \small velocity\\ \sf \small space}};
(8,-3)*{\color{red} y};
(18,-1)*{\color{red} \txt{\sf \small observer}};
(31,-7)*{\R^3_y};
(47,-5)*{\txt{\sf \small `space' from  $y$'s \\ \sf \small perspective}};
(50,0)*{};
(-50,0)*{};
\endxy
\]
An observer thus naturally splits spacetime fields into spatial and temporal parts. 
 
In more general spacetimes, this picture is valid only `infinitesimally,' on each tangent space.  A \define{field of observers} is a unit future-directed timelike vector field $u$---something any time-oriented Lorentzian manifold has.  Such a field suffices to split fields on a {\em background} spacetime into spatial and temporal parts.  But here we are interested in general relativity, where the metric---and hence the definition of observer---is to be determined by the dynamics.  Can we define `observers' without using the metric?

Fortunately, the \define{coframe field} in first-order gravity locally maps spacetime vectors to vectors in $\R^{3,1}$, which we view as an \define{internal spacetime}.   This lets us translate between an observer field and a more primitive notion: a field of \define{internal observers}, assigning an observer $y(x)\in H^3\subset \R^{3,1}$ to each point $x$ of spacetime. 

\begin{wrapfigure}{right}{4cm}
\vspace{0pt}
\xy  
(0,0)*{\includegraphics{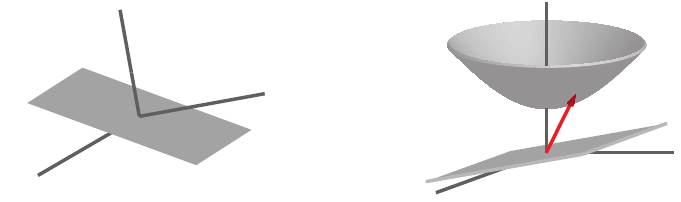}};
(-20,18)*{\text{\small \sf physical spacetime}};
(-20,14)*{\scriptstyle \text{\sf (one tangent space)}};
(20,18)*{\text{\small \sf internal spacetime}};
(-35,10)*{\scriptstyle T_xM};
(-22,-2)*{\scriptstyle x};
(28,0)*{\scriptstyle \color{red} y(x)};
(35,11)*{\scriptstyle \R^{3,1}};
(-23,-8)*{\scriptstyle \ker \hat u}="A";
"A";(-17,-4)**\dir{-};
(22,-10)*{\scriptstyle \R^3_y}="B";
"B";(18,-6)**\dir{-};
{\ar_{E}@/^.2pc/(-14,-3)*{};(12,-6)*{}}; 
(0,-17)*{\scriptstyle \text{\sf $E$ restricts to a `spatial coframe' $\ker \hat u \to \R^3_y$}};
(-10,-15);(-2,-8)**\dir{-};
\endxy
\vspace{-60pt}
\end{wrapfigure}
Starting with a smooth spacetime manifold $M$, besides the field $y$, we need:
\begin{itemize}
\item a nowhere-vanishing 1-form $\hat{u}$,
\item an $\R^3_y$-valued 1-form $E$  such that 

\rule{2cm}{0em}$e := E + \hat u\, y$

is a nondegenerate coframe field. 
\end{itemize}

\noindent 
The observer field $u$ itself is found by solving $y = e(u)$.  This implies $\hat u$ is dual to $u$, so from $u$'s perspective, `space' is the kernel of $\hat u$.  $E$ annihilates $u$ and so is `purely spatial.'  

All other differential forms similarly split into a \define{spatial part} annihilating $u$, and a \define{temporal part} of the form $\hat u\wedge X$.  In particular, the spin connection is given by
$$\omega=\Omega+\hat{u}\,\Xi$$ 
where the spatial 1-form $\Omega$ and the scalar $\Xi$ both live in $\mathfrak{so}(3,1)$. These constructions are clearly analogous to how spacetime fields are built in ADM gravity \cite{adm}. 

In this language, classical field equations split neatly into spatial equations constraining `initial values' of the fields and temporal equations corresponding to dynamics.  For example, the spatial part of the vacuum Einstein equation $e\we R =0$ is
\[
 E\we\left( \Rs[\Omega] + \Xi\, d^\perp \hat u \right) = 0
\]
where the \define{spatial differential} $d^{\perp}= d-\hat{u}\wedge\L_u$ depends on the Lie derivative $\L_u$ along $u$, and $\mathfrak{R}[\Omega]=d^{\perp}\Omega+\Omega\wedge\Omega$ is the \define{spatial curvature}. 

The resulting equations also follow from an action given in \cite{lorentz} with $\hat u$ as a background structure analogous to `$dt$' in foliation-based approaches; for $\hat{u}=N\,dt$, where $N$ is the lapse, our formulation reduces to ADM, rewritten in coframe variables.

But $\hat u$ need not be of the form $N\,dt$, even locally, in which case the distribution of spatial hyperplanes provided by $\hat u$ will not be tangent to any foliation:
\[ 
\xy
(0,0)*{\includegraphics[height=4.5cm]{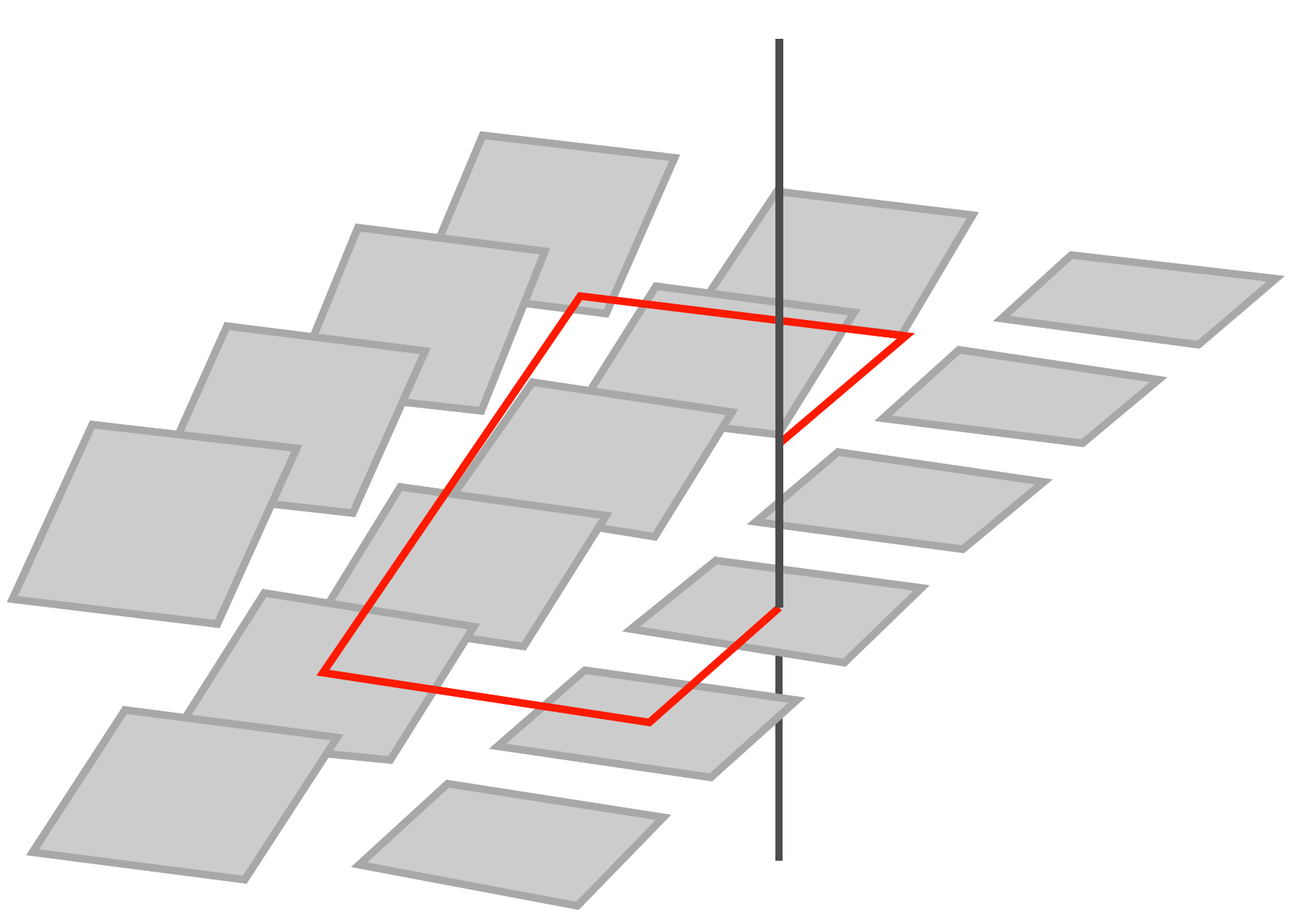}};
(-50,0)*{\txt{\small \sf spatial \\ \small \sf hyperplanes\\ \small ($\ker(\hat u)$)}}="A";
"A"+(10,3);(-20,4)**\dir{-};
"A"+(11,-2);(-25,-5)**\dir{-};
"A";(-24,-18)**\dir{-};
(50,0)*{\txt{\small \sf attempting a `loop  \\
      \small \sf in space' can result \\
      \small \sf in a time translation}}="B";
(15,-2)="C";
"B";"C"**\dir{-};
"C";(7.5,1)**\dir{-};
"C";(7.5,-7)**\dir{-};
\endxy
\]
The Frobenius theorem implies the local condition for $\hat u$ to determine a foliation is $d^{\perp}\hat{u}=0$, in which case our equations reduce to a more familiar form. 

But why allow configurations with $d^\perp \hat u \neq 0$? 

We have already mentioned that doing so avoids making global restrictions other than time-orientability.  More importantly, configurations with $d^\perp \hat u \neq 0$ are related to foliations by a {\em natural symmetry}.  Lorentz gauge transformations act on internal observers, $y\mapsto \Lambda\,y$, but because $y=e(u)$ one can transform either the coframe $e$ or the observer field $u$, giving two distinct types of local Lorentz symmetry.  Transforming $u$ will in general map a foliation to a configuration where $d^{\perp}\hat{u}\neq 0$; invariance under those transformations is the analog in our framework of independence of foliation in standard Hamiltonian approaches. General relativity {\em is} invariant because $\omega$ and $e$ are, but this symmetry is lost in some proposals for going beyond general relativity, as discussed later.

\section*{Lorentz covariance and Ashtekar variables}

An internal observer $y$ is fixed by a subgroup $\SO(3)_y\subset \SO(3,1)$, and Lorentz group representations split accordingly:
\[
\xymatrix@C=.2cm@R=3em{
& \!\!\so(3,1)\!\! \ar[dl]_{\pi_y} \ar[dr] & & \R^{3,1} \ar[dl] \ar[dr] & & \scriptstyle \leftarrow \text{ \sf adjoint and fundamental reps of $\SO(3,1)$} \\ 
\so(3)_y\hspace{-2em} & & \R^3_y & & \R^1_y & \scriptstyle \leftarrow \text{ \sf observer-dependent reps of $\SO(3)_y$}
}
\]
A local Lorentz transformation changes these splittings, but changes also the rotation group $\SO(3)_y\cong \SO(3)$, so all fields transform consistently. 

As an example, consider the $\so(3)_y$ part of the spatial connection, $\OMs = \pi_y\,\Omega$, 
where $\pi_y$ projects onto $\so(3)_y$.  Under a local Lorentz transformation of $\omega$, we get:
\[
y\mapsto y' = \Lambda\, y\,  \qquad
 \OMs \mapsto \OMs'=\Lambda^{-1}\,\OMs \,\Lambda + \pi_{y'} (\Lambda^{-1} \,d^{\perp}\Lambda)
\]
so that ${\bf \Omega'}$ lives in $\so(3)_{y'}$.  Similarly, the $\R^3_y$-valued `triad' $E$ transforms to take values in $\R^3_{y'}$. 



Under transformations living in $\SO(3)_y$, $\OMs$ and $E$ have just the right behavior for a spatial connection and triad.  One can show \cite{lorentz} that they generalize Ashtekar variables \cite{ashtekar}, in the real form due to Barbero \cite{barbero}.  In the Ashtekar-Barbero formulation, the apparent breaking of Lorentz symmetry down to $\SO(3)$ arises by fixing $y$ and therefore the subgroup $\SO(3)_y$ once and for all.


For us, this breaking occurs `spontaneously': at each spacetime point, $y\in H^3$ selects the subgroup $\SO(3)_y\subset \SO(3,1)$. By transforming $y$ along with the dynamical variables, the action of the full Lorentz group is maintained.

Breaking $\SO(3,1)$ symmetry spontaneously has two nice side-effects.  First, it side\-steps second class constraints that must be dealt with in related connection-based approaches.  Second, it makes the pair $(\OMs,E)$ into a `spatial Cartan connection,' making a precise link between `geometrodynamics' and `connection dynamics.'  See our papers \cite{lorentz,broken} for details.

\section*{Time evolution and covariant phase space}

In foliation-based approaches, `time evolution' is a particular 1-parameter family of spacetime diffeomorphisms: the flow generated by the vector field $\partial_t$, moving each spatial slice into the future by intervals of the arbitrary `time' function $t$:
\begin{center}
\begin{picture}(150,100)
\bezier{123}(5,0)(20,20)(10,50)\bezier{432}(10,50)(0,80)(10,100)
\bezier{123}(130,0)(135,20)(130,50)\bezier{432}(130,50)(120,80)(130,100)
\put(25,0){\vector(1,3){8}}\put(50,0){\vector(0,3){10}}\put(75,0){\vector(-1,3){4}}\put(100,0){\vector(1,3){7}}
\put(25,25){\vector(0,3){12}}\put(50,25){\vector(1,3){6}}\put(75,25){\vector(2,3){8}}\put(100,25){\vector(2,3){12}}\put(125,25){\vector(0,3){17}}
\put(25,50){\vector(-1,3){4}}\put(35,50){\vector(0,1){10}}\put(50,50){\vector(1,1){12}}\put(75,50){\vector(1,1){17}}\put(100,50){\vector(1,3){7}}
\put(25,75){\vector(1,3){4}}\put(50,75){\vector(2,3){12}}\put(75,75){\vector(1,3){7}}\put(100,75){\vector(2,3){13}}
\put(135,20){$t=t_0$}\put(133,35){$t=t_0+1$}\put(128,65){$t=t_0+2$}
\color{blue}\put(14,25){\line(1,0){118}}\bezier{453}(14,32)(25,39)(56,43)\bezier{711}(56,43)(70,43)(83,37)\bezier{812}(83,37)(95,34)(112,43)\bezier{240}(112,43)(118,47)(125,42)\bezier{412}(125,42)(127,40)(132,36)
\bezier{453}(13,40)(25,49)(56,55)\bezier{711}(56,55)(70,55)(83,59)\bezier{812}(83,59)(95,63)(112,62)\bezier{240}(112,62)(120,62)(126,67)
\end{picture}
\end{center} 

An observer field $u$ also generates a flow representing `time evolution'. Since $u$ is normalized, this flow is parameterized by {\em proper time} of the observer field:
\begin{center}
\begin{picture}(150,100)
\bezier{123}(5,0)(20,20)(10,50)\bezier{432}(10,50)(0,80)(10,100)
\bezier{123}(130,0)(135,20)(130,50)\bezier{432}(130,50)(120,80)(130,100)
\put(25,0){\vector(1,3){5}}\put(50,0){\vector(0,3){14}}\put(75,0){\vector(-1,3){5}}\put(100,0){\vector(1,3){5}}
\put(25,25){\vector(0,3){14}}\put(50,25){\vector(1,3){5}}\put(75,25){\vector(2,3){13}}\put(100,25){\vector(2,3){13}}\put(125,25){\vector(0,3){15}}
\put(25,50){\vector(-1,3){5}}\put(35,50){\vector(0,1){14}}\put(50,50){\vector(2,3){13}}\put(75,50){\vector(2,3){13}}\put(100,50){\vector(1,3){5}}
\put(25,75){\vector(1,3){5}}\put(50,75){\vector(2,3){13}}\put(75,75){\vector(1,3){5}}\put(100,75){\vector(2,3){13}}
\color{red}{\bezier{235}(20,0)(27,25)(22,50)\bezier{471}(22,50)(15,75)(20,100)}
\bezier{317}(30,0)(40,40)(50,50)\bezier{580}(50,50)(55,58)(67,72)\bezier{210}(67,72)(70,74)(80,100)
\bezier{432}(52,0)(55,30)(72,50)\bezier{245}(72,50)(80,60)(95,75)\bezier{671}(95,75)(100,81)(108,100)
\bezier{700}(85,0)(90,25)(105,50)\bezier{625}(105,50)(118,70)(125,100)
\end{picture}
\end{center}

Thus, while we have no global notion of space, there is a canonical way to push the {\em whole spacetime} forward by one second of proper time of the observers.   


But how do we define {\em phase space} without a foliation? 

The \define{covariant phase space} of general relativity is its space of solutions, a natural covariant generalization of the `canonical' phase space \cite{crnkovic,gotay}.  However, not dividing spacetime into space and time, it lacks any obvious link to the conceptual picture of spatial configurations changing in time.

The observer-based formulation could provide this link. On one hand, if we choose an observer field corresponding to a foliation, we recover canonical gravity. On the other, everything transforms covariantly under change of observer, a local gauge choice. Adjoining the observer field $y$ gives us a covariant phase space in which spatial and temporal variables are clearly distinguished, without spoiling local Lorentz symmetry.

\section*{The space of observers}

In general relativity, just as there is no canonical spacelike foliation, there is no canonical choice of observer.  Faced with such a situation, rather than making an arbitrary choice, one can {\em simultaneously consider all possible choices.}  Individual choices are arbitrary; the {\em space} of choices is canonical.  

In foliation-based approaches, this philosophy is not very helpful: the `space of all spacelike foliations' is too unwieldy, and hard to interpret physically.

On the other hand, \define{observer space}, the space of all possible observers, has manifest physical meaning, and simple topology: it is a 7-dimensional manifold isomorphic to the `unit future tangent bundle' of spacetime, locally a product of spacetime with velocity space $H^3$.  In \cite{wip} we reformulate general relativity directly on observer space, essentially by pulling fields back along the natural projection 
\[
   \text{observer space } \longrightarrow \text{ spacetime}\,.
\]
A connection pulled back to observer space will be flat in the `velocity' directions, reflecting the symmetry under changes of observer.  

General relativity respects this symmetry, but does {\em nature}?  As we and all our instruments are `observers,' we cannot probe spacetime geometry directly in any observer-independent way.  The empirical evidence for symmetry under a change of observer could be challenged by future observations.

Several modifications of general relativity currently of interest can be studied using observer space. First, there is a growing interest in models that do not treat spacetime isotropically \cite{cdt,shape,horava}. Since points in observer space correspond to directions in spacetime, these anisotropic theories might be described very naturally in these terms.

Perhaps more compelling is the question of whether spacetime itself plays any fundamental role in physics.  Once we have lifted the theory to observer space, do we still have any need for spacetime?  In fact, starting with observer space, we can {\em reconstruct spacetime}---but only when certain `integrability conditions' hold \cite{wip}.   The `relative locality' proposal \cite{relative} suggests the notion of spacetime itself may be {\em observer-dependent}.  Observer space provides a natural perspective from which to study this possibility, with the potential to move beyond `special' and on to `general' relative locality.

\end{document}